\documentstyle[12pt]{article}
\begin{document}
\vspace*{-1.5cm}
{\obeylines
July 1995 \hfill PAR--LPTHE 95-38\\[-5mm]
\hfill ULB--TH 10/95 \\[-5mm]
\flushright hep-th/9507061
}
\vskip 1cm
\centerline{\bf BLACK HOLE ENTROPY AND STRING INSTANTONS }
\vskip1cm
\begin{center} Fran\c{c}ois~Englert \footnote{ e-mail:
fenglert@ulb.ac.be}\\[.3cm] {\it Service de Physique Th\'eorique\\
Universit\'e Libre de Bruxelles, Campus Plaine, C.P. 225\\
Boulevard du Triomphe, B-1050 Bruxelles, Belgium }
 \vskip .5cm
Laurent Houart \footnote{Chercheur associ\'e CNRS, e-mail:
lhouart@lpthe.jussieu.fr}  and \ Paul~Windey \footnote{e-mail:
windey@lpthe.jussieu.fr}\\[.3cm] {\it Laboratoire de Physique
Th\'eorique et Hautes Energies \footnote{ Laboratoire associ\'e No.
280 au CNRS.}\\  Universit\'e Pierre et Marie Curie, Paris VI\\ Bte
126, 4 Place Jussieu, 75252 Paris cedex 05, France} \end{center}
\vskip 1cm \begin{abstract} \noindent  The coupling of a string to
gravity allows for Schwarzschild black holes whose entropy to area
relation is $S=(A/4)(1-4\mu)$, where $\mu$ is the string tension.
This departure from the A/4 universality  results from a string
instanton generating a   black hole with smaller entropy at a
temperature exceeding the Hawking value. The temperature is
sensitive to  the presence of classical matter outside the black
hole horizon but the entropy is not. The horizon materializes at
the  quantum level. It is conjectured that such a macroscopic non
local effect may be operative in retrieving information from
incipient black holes.
\end{abstract}

\section{Introduction} \label{intro}

In absence of back reaction, an incipient black hole of mass  $M$
emits a thermal flux at the Hawking temperature $\beta_H^{-1}$
given by \cite{haw1} \begin{equation} \label{mass} \beta_H = 8\pi
M.   \end{equation}  The flux  is filtered by the centrifugal
barrier and consists predominantly of low angular momentum  quanta
which have leaked out  and escape towards infinity.

To study the thermodynamic properties of the black hole, one may
consider, instead of  its irreversible evaporation,  reversible
exchanges between the black hole and surrounding matter in a state
of thermal equilibrium. Assuming that $M$ is equal to its
thermodynamic energy $E$, one may characterize such exchanges by
\begin{equation} \label{bdmf}
 dS = \beta_H dM    \end{equation}  where $S$ is the black hole
entropy. This gives, up to a constant of integration, the
Bekenstein-Hawking \cite{bek} area entropy  namely
\begin{equation}  \label{area}
 S = {A\over 4}    \end{equation} where $A=16\pi M^2$ is the area
of the event horizon.

The entropy Eq.\ (\ref{area}) is independent of the   particular
collapse and is therefore expected to be a genuine property of
the   Schwarzschild  background geometry.  To display  this property, it is
convenient to extend maximally the original Schwarzschild  metric. The
extended metric describes     an ``eternal'' black hole which
is time reversal symmetric and is therefore suitable for describing
thermal equilibrium. This extension introduces, in addition to the
dynamical regions separated from the Schwarzschild  static patch by future
and past horizons, a copy of the original Schwarzschild  patch.  When
referring to a Schwarzschild  background, we shall always mean the original
patch.

To understand how entropy can hide in the geometry, we first recall
that to compute thermal correlation functions and partition
functions in field theory in flat Minkowski space-time one can use
path integrals in periodic imaginary time.  The period $\beta$ is
the inverse temperature and can be chosen freely. This method was
generalized to compute matter correlation functions in static
curved backgrounds.  For the Schwarzschild background, the analytic
continuation to imaginary time defines an Euclidean manifold
everywhere except at the analytic continuation of the horizon,
namely the 2-sphere at $r=2M$, $M$ being the black hole  mass. This
ambiguity is usually removed by assuming that the Euclidean
manifold is regular at $r=2M$. The Euclidean black hole  obtained
in this way has a periodicity $\beta$ uniquely defined in terms of
the black hole  mass. The relation is $\beta = 8\pi M$ and gives
back the temperature of the Hawking radiation Eq.\ (\ref{mass}).

Gibbons and Hawking \cite{gh1}   extended the analytic continuation
to the gravitational action, restricting the hitherto ill-defined
path integral over metrics to a saddle point in the Euclidean
section. To constitute such a saddle the Euclidean black hole must
be regular given that a singularity at $r=2M$ would invalidate the
solution of the Euclidean Einstein equations. The partition
function evaluated on this saddle is interpreted as $e^{-\beta F}$,
where $\beta$ is equal to $\beta_H$ and $F$ the free energy of the
background geometry. This yields back the Bekenstein-Hawking  area
entropy  Eq.\ (\ref{area}).

The Gibbons-Hawking procedure gives the expected thermodynamic
results but its justification is not obvious. Were it not for the
saddle condition, the functional integral over metrics would
formally amount to a computation of a partition function. However,
not only is the Euclidean functional integral over all metrics
meaningless but the partition function of the canonical  ensemble
for a black hole whose density of states would be given by the
exponential of the entropy Eq.\ (\ref {area}) does not exist
because of the divergence at large masses. Clearly, this divergence
is cut off by the saddle  and we shall give a thermodynamic
argument to justify  this procedure.

The saddle prescription can be used to show that the  relation Eq.\
(\ref{mass}) between the temperature and the black hole mass is
affected by classical surrounding matter but that the entropy
remains unchanged and is still given by Eq.\ (\ref{area}). This
value of the entropy seems therefore to depend only on the black
hole mass. This is not the case. A different relation between
entropy and area emerges when a conical singularity is present in
the Euclidean section at $r=2M$ \cite{ehw}.

Many authors \cite{cpw,dgt,btz,tei,ct,su} have introduced a conical
singularity at $r=2M$. This modifies the Euclidean periodicity of
the black hole and therefore the temperature. However if the source
producing this singularity is not taken into account, the Euclidean
black hole solution is not a saddle point of the functional
integral and such a procedure cannot be used to find the
thermodynamic properties of the background.  A true Euclidean
saddle point can nonetheless be maintained by introducing a
string term in the action.  The string is  treated in first
quantization. One may interpret this as a fundamental string
term or alternatively as a thin vortex limit of
a Nielsen-Olesen flux tube in the spontaneously broken phase of a
field theory. The  conical singularity arises from a
string ``instanton'' wrapping   around the Euclidean continuation
of the horizon and the associated deficit angle is determined by
the string tension.  The temperature   varies continuously above
the Hawking value as a function of the string tension but, as shown
in reference \cite{ehw}, it now necessarily entails a variation of
the entropy versus area ratio: this ratio takes value in the
interval $[0,1/4]$, the lower limit being approached when the cone
degenerates.

In this paper we    generalize   this result and discuss  its
significance. In Section  \ref{stringinstanton} we review for
completeness the results of reference \cite{ehw}. We introduce by
hand a Nambu-Goto string in the action and require, as a working
hypothesis ---to be justified in the   Section
\ref{euclideansaddle}--- that the Schwarzschild black hole
solutions analytically continued to imaginary time satisfy the
Euclidean Einstein equations. This requirement fixes uniquely the
Euclidean periodicity  in terms of the string tension for a black
hole of mass $M$ in the presence of a Euclidean string
``instanton'' wrapped once around the horizon.  We then give the
modification of the relation between entropy and area resulting
from the change in periodicity induced by the string. This comes
about because the product of the temperature and the entropy, a
purely classical quantity, retains its traditional Schwarzschild
value independently of the string tension. In Section
\ref{equationofstate}  we show that,  at fixed string tension,
the   relation   between entropy and area in presence of the string
instanton is, in contradistinction with the temperature,
insensitive  to the introduction of classical surrounding matter.
In Section  \ref{euclideansaddle} a thermodynamic argument is
presented to justify the working hypothesis used throughout  the
article. It leads to the necessary introduction of the string if
the Euclidean periodicity is chosen to be different from its
conventional value.  It also fixes the value of the string tension
to be given by the Euclidean saddle constraint. We then confront
the problem posed by the existence of several Euclidean saddles,
namely those arising from configurations with and  without a string
wrapping once, and possibly many times, around the 2-sphere $r=2M$
in the Euclidean section. We  conclude that while the no instanton
solution defines a stable  black hole with minimal temperature and
maximal entropy $A/4$, the other solutions  correspond to
thermodynamically distinct overheated black holes with lower
entropy.  In   Section \ref{oneloop}  we show,  in a
simplified model, that the     string instanton induces on the
horizon, in the Lorentzian background,  singularities   in
the energy-momentum tensor of quantum fields at the loop level.
This amounts to a materialization of the horizon at the quantum
level. We argue that a similar situation may arise for incipient
black holes and we suggest that such a non local quantum effect may
play a crucial role in retrieving the information hidden by, or
stored on the horizon.

\section{The String Instanton} \label{stringinstanton}

The Lorentzian action for gravity coupled to matter fields is taken
to be \begin{equation} \label{L-action} I={1\over 16\pi}\int_M\
\sqrt{\vert g\vert} R -{1\over 8\pi}\int_{\partial M}\ \sqrt{\vert
h\vert}K + I_{matter}. \end{equation} Here ${1\over 16\pi}\int_M\
\sqrt{\vert g\vert}R$ is the usual Einstein-Hilbert action, $K$ is
the trace of the extrinsic curvature on the boundary $\partial M$
of the four dimensional manifold $M$, and $h$ the determinant of
the induced metric.

The introduction of the $K$-term requires explanation. We will
justify it briefly and refer the interested reader to the recent
detailed discussion of Hawking and Horowitz \cite{haho}. The
Einstein-Hilbert action contains second-order derivatives of the
metric. If the system evolves between two non intersecting
spacelike hypersurfaces these second derivative terms can  be
transformed by partial integration into boundary terms on these
spacelike surfaces and on  timelike surfaces. Explicitly these
boundary terms stem from the integral of the four-divergence
$\partial_{\mu} {\omega}^\mu$ where \begin{equation} \label{omega}
\omega^\mu = -{1 \over 16\pi}\left( \partial_\nu( \sqrt{\vert
g\vert} g^{\mu \nu})+ g^{\mu \nu} \partial_\nu \sqrt{\vert
g\vert}\right).  \end{equation}  Their contribution to the action
Eq.\ (\ref {L-action}) is cancelled by the $K$-term. The absence of
boundary terms on the spacelike surfaces is necessary for the
consistency of the Hamiltonian formalism. However, for the
asymptotically flat spaces considered here, the $K$-term introduces
divergences at spacelike infinity.  These can be removed by
subtracting from Eq.\ (\ref {L-action}) a $K$-term at infinity in
flat space. It can then be verified that the subtracted action
\begin{eqnarray} \nonumber I-I_0&=&{1\over 16\pi}\int_M\
\sqrt{\vert g\vert} R -{1\over 8\pi}\int_{\partial M}\ \sqrt{\vert
h\vert}K \\ \label{subtraction} &&+ {1\over 8\pi}\int_{(\partial
M)_\infty}\ \sqrt{\vert h_0\vert} K_0 + I_{matter} \end{eqnarray}
yields the correct ADM mass as the on-shell value of the
Hamiltonian. The action Eq.\ (\ref {subtraction}) can now be
written as a Hamiltonian action and the path integral over metrics
can be formally defined.

We will consider a system where the matter is an elementary
Nambu-Goto string. Its action is given by \begin{equation}
\label{ng}  I_{matter}\equiv I_{string}= -\mu \int\ d^2\sigma
\sqrt{\vert\gamma\vert}, \end{equation}  where $\mu$ is the string
tension and $\gamma$ is the determinant of the induced metric on the
worldsheet:  \begin{equation}
\gamma_{ab}(z)=g_{\mu\nu}(z)\partial_az^\mu\partial_b z^\nu.
\end{equation} In the presence of a string, the Lorentzian Einstein
equations still admit ordinary black hole solutions corresponding
to trivial string solutions spanning zero area. The continuation of
these solutions to imaginary time is given by the metric
\begin{equation} \label{schw} ds^2=\left(1-{2M \over r}\right)
dt^2+ \left(1-{2M \over r}\right)^{-1} dr^2+r^2 d\Omega^2
\end{equation} for $r > 2M$. To determine the Euclidean manifold at
$r=2M$ we impose the Euclidean Einstein equations. As discussed in
the Introduction, the free energy will then be computed on the
saddle of the Euclidean action.

The action Eq.\ (\ref {L-action}) with matter term Eq.\ (\ref {ng})
can be extended to Euclidean metrics.  The subtracted Euclidean
action reads  \begin{eqnarray} \nonumber \widetilde I-\widetilde
I_0&=&-{1\over 16\pi}\int_{\widetilde M}\ \sqrt{ g } R +{1\over
8\pi}\int_{\partial \widetilde M}\ \sqrt{ h }K \\ \label{action}
&&- {1\over 8\pi}\int_{(\partial \widetilde M)_\infty}\ \sqrt{ h_0
} K_0 + \mu \int\ d^2\sigma \sqrt{\gamma}. \end{eqnarray}  Here the
world sheet has the topology of a 2-sphere. Because the Euclidean
black holes Eq.\ (\ref{schw}) have only one boundary, namely at
infinity, we must take $\partial \widetilde M = (\partial
\widetilde M)_\infty$ in the $K$-term. The $K_0$-term subtraction
has to be performed with the extrinsic curvature in flat Euclidean
space. The independent variables in Eq.\ (\ref{action}) are the
components of the metric $g_{\mu\nu}$ and the string coordinates
$z^\mu$. The variation of the action with respect to $g_{\mu\nu}$
gives the Euclidean Einstein equations: \begin{equation}
\label{einstein} R_{\mu\nu}(x)-{1 \over 2} g_{\mu\nu}(x)\ R(x)=8
\pi T_{\mu\nu} (x) \end{equation}  where  \begin{equation}
T^{\mu\nu}(x)=-\mu\int d^2\sigma
\sqrt{\gamma}\gamma^{ab}\partial_az^\mu  \partial_b z^\nu {1 \over
\sqrt{ g(z)}} \delta^4 (x-z). \end{equation}  Variations with
respect to $z^{\mu}$ give rise to the stationary area condition for
the string.

The Einstein equations Eq. (\ref {einstein}) admit ordinary
Euclidean black holes solutions corresponding to trivial zero
string area.  However the topology of the Euclidean manifold
allows for a non-trivial string solution with minimal area which
wraps around the continuation of the horizon $r=2M$.   All
solutions are correctly described by the metric Eq. (\ref{schw})
but the non-trivial one has a curvature singularity at $r=2M$.   To
analyze its structure, we write the metric in the vicinity of the
horizon in terms of the proper radial length $\xi$ at fixed time
$t$ \begin{equation} \label{radial} \left(1-{2M\over
r}\right)^{-{1\over 2}} dr = d\xi\qquad \xi=0 \quad \mbox{at} \quad
r=2M, \end{equation} which gives to order $\xi^2$ \begin{equation}
\label{sxi} ds^2 = \xi^2 \left({dt\over 4 M}\right)^2 +  d\xi^2
+r^2 d\Omega^2. \end{equation} The Euclidean space is regular at
$r=2M$ when the periodicity of $t$ is \begin{equation}
\label{period}  \beta_H=8\pi M \end{equation} and hence when it
coincides with the inverse Hawking temperature $\beta_H^{-1}$ of
the thermal flux emitted by the incipient black hole Eq.
(\ref{mass}). This periodicity is modified by the presence of a
string wrapped around the horizon. The trace of Einstein equations
Eq. (\ref{einstein}) gives \begin{equation} \label{trace}
\int_{\widetilde M} \sqrt{  g } R = 16\pi \mu A, \end{equation}
where $A$, the area of the string, is equal to the area of the two
sphere at $r=2M$.  The entire contribution to the integral comes
from the string singularity. To evaluate $\int \sqrt{g} R$ when $R$
is zero everywhere except on the horizon, one can consider a
tubular neighborhood $S^2 \times D_\epsilon$ of $r=2M$, of radial
length $\epsilon$, and let $\epsilon$ tend to zero\cite{btz,tei}.
This gives \begin{equation} \label{r4asr2} \int_{S^2 \times
D_\epsilon} \sqrt{g} R = A\int_{D_\epsilon} \sqrt{ ^{(2)}g}\,
\,^{(2)}\!R. \end{equation} The Euler characteristic  of the disc
topology is $\chi=1$. It is given by \begin{equation} \label{euler}
\chi={1\over 4\pi}\int_{D_\epsilon}\ \sqrt{ ^{(2)}g} \, ^{(2)}\!R
-{1\over 2\pi} \oint \sqrt{h} K. \end{equation} For a curvature
singularity at the origin with \begin{equation} {1\over
4\pi}\int_{D_\epsilon}\  \sqrt{ ^{(2)}g} \, ^{(2)}\!R = \eta,
\end{equation} we find, from Eq. (\ref{euler}),  \begin{equation}
-\oint \sqrt{h} K=2\pi(1-\eta). \end{equation} This signals the
existence of a conical singularity with deficit angle $2\pi \eta$.
Comparing Eq. (\ref{trace}) and Eq.  (\ref{r4asr2}), we then
conclude that the string gives rise to a deficit angle
\begin{equation} \label{deficit} \eta=4 \mu. \end{equation} This is
the sole effect of the string instanton. From the explicit form of
the Schwarzschild solution given in Eq. (\ref{sxi}), we see that
the periodicity of $t$ has become: \begin{equation} \label{temp}
\beta=\beta_H \ (1-4 \mu). \end{equation} The string instanton has
raised the global temperature from $\beta^{-1}_H$ to $\beta^{-1}$.

We now evaluate the free energy of the black hole. The contribution
of the string term to the action Eq.\ (\ref{action}) exactly
cancels the contribution of the Einstein term as seen from Eq.\
(\ref{trace}). The boundary terms at asymptotically large
$r=r_{\infty}$ are thus the only ones contributing to $\beta F$.
Using Eqs.\ (\ref{omega}) and (\ref{schw}) we find \begin{equation}
\label{kterm} {1\over 8\pi}\int_{\partial \widetilde M=(\partial
\widetilde M)_\infty}\ \sqrt{h}K=-\beta
\left(r_{\infty}\left(1-{2M \over r_{\infty}}\right) +{M\over
2}\right).   \end{equation} The subtracted term is computed
similarly in the flat metric \begin{equation} \label{flat} ds^2 =
\left(1-{2M \over r_{\infty}}\right) dt^2 + dr^2+ r^2 d\Omega^2,
\end{equation} where $t$ has the periodicity $\beta$ given by Eq.\
(\ref{temp}). The subtraction term is  \begin{equation}
\label{kflat} {1\over 8\pi}\int_{(\partial \widetilde M)_\infty}\
\sqrt{h_0}K_0=-\beta r_{\infty} \left(1-{2M \over
r_\infty}\right)^{1 \over 2}. \end{equation} The free energy is
given by  \begin{equation} \label{free} F(\beta,\mu) = \beta^{-1}
(\widetilde I-\widetilde I_0)_{saddle} = {M \over 2}. \end{equation}
Note that the free energy has the same value it had in the absence
of the string instanton. However the thermodynamic potential $F$ is
now a function of $\beta$ and the external parameter $\mu$.  The
entropy is given by:  \begin{equation} \label{defe}  S = \beta^2
\left. {\partial F \over \partial \beta} \right|_\mu ={\beta^2
\over 2} \left. {\partial M \over \partial \beta} \right|_\mu=
{\beta^2 \over 16 \pi (1-4\mu)}. \end{equation} Using Eq.\
(\ref{temp}) we find \begin{equation} \label{entropy} S=(1-4 \mu)
{A \over 4}. \end{equation} This was the central result of
reference \cite{ehw}. The introduction of the external parameter
$\mu$ has enabled us to define a black hole with fixed mass $M$ at
a temperature other than the usual $\beta_H^{-1}$. When the
temperature is not the Hawking temperature, the entropy changes
from $A/4$ to Eq.\ (\ref{entropy}). It follows from Eq.\
(\ref{temp}) and Eq.\ (\ref{entropy}) that the product of the
entropy and the temperature is constant for a given mass,
independent of the string tension \begin{equation} \label{state}
\beta^{-1} S= \beta^{-1}_H S_H = {M\over 2}, \end{equation} where
$S_H=A/4$.

To complete the thermodynamic analysis, we first verify using
Eqs.(\ref{state})  and (\ref{free}) that the energy of the solution
is unchanged by the presence of the string instanton:
\begin{equation} \label{energy} E= F + \beta^{-1} S ={M \over 2}
+\beta_H^{-1}{A \over 4}=M. \end{equation} We immediately conclude
from this equation  that the  identity \begin{eqnarray} \label{dmf0}
dM &=&  \beta_H^{-1} dS_H \\ \label{dfdmu}
   &=& \beta^{-1} dS +\beta^{-1}A d\mu, \end{eqnarray} is nothing
but an expression of the first principle of black hole
thermodynamics in the presence of a string.

The area entropy of an eternal black hole can thus vary from the
standard value $A/4$ to $0$ as a function of the string tension.
All these black holes differ only by the instanton effect. One
cannot distinguish them by their mass or even by their Lorentzian
metrics. They differ through quantum effects, not in classical
quantities such as $M/2$, the product of $\beta^{-1}$ and $S$.
Indeed the temperature is proportional to $\hbar$ while the entropy
is inversely proportional to $\hbar$. In this sense the instanton
provides a quantum hair \cite{cpw,dgt} which affects the
expectation values of operators outside the horizon. Quantum
effects will be further analyzed in Section \ref{oneloop}.

The above result rely crucially on the working hypothesis that the
free energy of the Schwarzschild  background is obtained from the saddle of
the Euclidean functional integral at   analytically continued
solutions of the Einstein equation defining the Lorentzian
background. We have also assumed   that continuations corresponding
to different instantons, and in particular to the trivial one with
$\beta =\beta_H$, generate distinct black holes. The discussion of
the validity of these assumptions will be the object of Section
\ref{euclideansaddle}. Before confronting the issue however, we
shall establish the general nature of the entropy Eq. (\ref
{entropy}) emerging from this working hypothesis.

\section{The Black Hole in Presence of Classical Matter}
\label{equationofstate} We now show that the entropy Eq.\ (\ref
{entropy}) is a genuine property of the horizon. Indeed we shall
prove that this result is not affected by classical static matter
outside the horizon. Classical  surrounding matter does change the
temperature but does not affect the entropy.

We surround the eternal black hole by a static and regular
distribution of matter which is a spherically symmetric  solution
of Einstein equations for a coupled gravity-matter system. The new
matter action in Eq.\ (\ref {L-action}) now reads\footnote{The
possible instabilities in the static matter distribution can always
be compensated by external forces.} \begin{equation}
\label{mattertotal} I_{matter} = I_{string} + I'_{matter}.
\end{equation} For $r>2M$, the Euclidean metric solution of the
Euclidean Einstein equations is given by \begin{equation}
\label{metric} ds^2=g_{tt}dt^2 +g_{rr} dr^2 +r^2 d\Omega^2
\end{equation} with \begin{eqnarray} \nonumber \label{time}
g_{tt}(r) &=& \left( 1-{2M(r)\over r}\right) \exp \left[- 8 \pi
\int_r^\infty  (-T^t_t(r') + T^r_r(r')) r'g_{rr}(r') \,dr' \right]\\
\label{space} g_{rr}(r)&= &\left(1-{2M(r)\over r}\right)^{-1}
\end{eqnarray} where \begin{equation}
 M(r)= M + \int_{2M}^r 4\pi r'^2(- T^t_t(r'))dr', \end{equation}
and $M$ is still the black hole mass. The regularity condition
characterizing the matter is $-T^t_t(2M) + T^r_r(2M)=0$. We
parametrize again the metric in the vicinity of the horizon in
terms of the proper length $\xi$ at fixed time $t$. Noticing that
$g_{tt}=O(\xi^2)$ we find \begin{equation} \label{sexi} ds^2 \simeq
\xi^2 \left({dt\over \lambda(0)}\right)^2 +  d\xi^2 +r^2 d\Omega^2
\end{equation} where in general   \begin{equation} \label{deflambda}
\lambda(\xi)\equiv {d g_{tt}^{1\over 2} \over d\xi}=
{1\over2}(g_{tt}g_{rr})^{-{1 \over 2}}{d g_{tt} \over  dr}.
\end{equation} This equation replaces Eq.\ (\ref {sxi}) in the
presence of matter.  The deficit angle engendered by the string is
still given by $2\pi\eta=8\pi\mu$ since the matter distribution,
which is regular, affects Eq.\ (\ref {r4asr2}) only to order
$\epsilon^2$ and hence vanishes in the limit $\epsilon\to 0$. The
inverse temperature of the black hole surrounded by regular static
matter thus becomes  \begin{equation} \label{tempmatter} \beta=
\beta_{Hm} (1-4\mu), \end{equation} where  \begin{equation}
\label{matemp} \beta_{Hm}  =  {2\pi\over \lambda(0)}
\end{equation} is the Hawking inverse temperature in the presence
of  matter (see for instance references \cite{ce,ce1,vis}).

We now compute the free energy from the Euclidean action. The new
subtracted Euclidean action can be evaluated on the solution of the
equations of motion   \begin{eqnarray} \label{fullaction}
\widetilde I'-\widetilde I'_0&=&-{1\over 16\pi}\int_{\widetilde
M_\epsilon}\
 \sqrt{ g }R+\mu \int\ d^2\sigma \sqrt{\gamma}\nonumber\\ &
&-{1\over 16\pi}\int_{\widetilde M\backslash \widetilde M_\epsilon
}\ \sqrt{ g } R + \widetilde I'_{matter} \nonumber \\ & & +{1\over
8\pi}\int_{(\partial \widetilde M)_\infty}\ \left(\sqrt{h}K
-\sqrt{h_0}K_0\right) \end {eqnarray} where $\widetilde M$ denotes
the Euclidean manifold of the solution and
 $ \widetilde M_\epsilon=S^2 \times D_\epsilon $ is a tubular
neighborhood of the horizon of length $\epsilon$. The first two
terms in Eq.\ (\ref{fullaction}) cancel each other as $\epsilon \to
0$ while $\widetilde I'_{matter}$ is equal for this static solution
to $\beta H'_{matter}$ where $ H'_{matter}$ is the Hamiltonian
constructed from $I'_{matter}$.  The surface terms \begin{equation}
{1\over 8\pi}\int_{(\partial \widetilde M)_\infty}\ \left(\sqrt{h}K
-\sqrt{h_0}K_0\right) = \beta {M_{tot} \over 2}, \end{equation} are
obtained by substituting in Eqs.\ (\ref {kterm}) and (\ref{kflat})
$M$ by $M_{tot}=M(r_\infty)$.    Thus Eq.\ (\ref {fullaction})
yields \begin{equation} \label {matterfree} F = -{1\over
16\pi}\int_{(\widetilde M\backslash \widetilde M_\epsilon )_t}
\sqrt{ g } R + H'_{matter}+ { M_{tot}\over 2} \end{equation} where
the $t$-subscripts indicate that the integrals are carried over the
three dimensional static domain obtained by intersecting
$\widetilde M\backslash \widetilde M _\epsilon$ by a constant
$t$-plane.  In order to obtain a simpler expression for $F$, we
start from the  Hamiltonian constraint, integrated over
$(\widetilde M\backslash \widetilde M_\epsilon)_t $
\begin{equation} -{1\over 16\pi}\int_{(\widetilde M\backslash
\widetilde M_\epsilon )_t} \sqrt{ g }\; ^{(3)}\!R + H'_{matter}=0
\end{equation} where $ ^{(3)}\!R$ is the three dimensional
curvature of $(\widetilde M\backslash \widetilde M_\epsilon)_t$.
Expressing $ ^{(3)}\!R $ in terms of $R$, \begin{equation}
\label{r3asr4} {1\over 16\pi}\int_{(\widetilde M\backslash
\widetilde M_\epsilon)_t} \left(\sqrt{g}R-\sqrt{g}\;
^{(3)}\!R\right)=-{1\over 8\pi}\int_{(\widetilde M\backslash  \widetilde
M_\epsilon )_t}
 \partial_i\left(\sqrt{ ^{(3)}\!{g}}g^{ij}\partial_j g_{tt}^{1 \over
2}\right) \end{equation} where $g_{ij}$ is the metric tensor of
the three surface, we recover   the integral mass
formula \cite{sma,bch,wal,ce,ce1} \begin{equation}
\label{smarr} -{1\over 16\pi}\int_{(\widetilde M\backslash
\widetilde M_\epsilon )_t}\ \sqrt{ g } R + H'_{matter} +
\beta_{Hm}^{-1} {A\over 4} - { M_{tot}\over 2}=0. \end{equation}
  The free energy is then obtained by comparing
Eqs.\ (\ref {matterfree}) and (\ref{smarr}): \begin{equation}
\label{fullfree} F= M_{tot} - \beta_{Hm}^{-1} {A\over 4}.
\end{equation} To compute the entropy, we need to take the
variation of $F$, hence the variation of $M_{tot}$.  Our starting
point will be the integral mass formula Eq.\ (\ref{smarr}). Notice
that its form is identical in the Euclidean or Lorentzian
formulation. In the latter case the integral is carried over a
three dimensional static domain which is the same as in the
Euclidean case. We thus skip the ``tilde'' notation and consider
the formula in the limit $\epsilon\to 0$  \begin{equation}
\label{smarr2} \lim_{\epsilon \rightarrow 0} \left( -{1\over
16\pi}\int_{(M \backslash M_\epsilon )_t}\ \sqrt{ |g| } R \right) +
H^\prime_{matter} + \beta_{Hm}^{-1} {A\over 4} - { M_{tot}\over
2}=0.   \end{equation} We now vary Eq.\ (\ref{smarr2}) along the
space of spherically symmetric static solutions \cite{ce,ce1}.  We
multiply Eq.\ (\ref{smarr2}) by a finite time interval $\Delta t$
and replace $\Delta t H^\prime_{matter}$ by $-I^\prime_{matter}$.
The variation of the two first terms in the fixed finite time
interval vanishes on the space of solutions except for the boundary
terms of the Einstein-Hilbert action and for the variation of
external non-gravitational parameters in the matter action. The
former is: \begin{eqnarray} \nonumber {1 \over \Delta t} \delta
\left( {1\over 16\pi}\int_{M}  \sqrt{ |g| } R \right)\! &=& \!
{1\over 16\pi \Delta t}\int_{M} \left( \partial_\eta (\sqrt{|g|}
g^{\mu \nu} \delta \Gamma^\eta_{\mu \nu})-\partial_\nu (\sqrt{|g|}
g^{\mu \nu} \delta \Gamma^\eta_{\mu \eta}) \right)\\ \nonumber & =
& {1 \over 2} (2M)^2 \delta \lambda (0) - \left[{1 \over 2} r^2
\delta \lambda + r (- g_{tt})^{1 \over 2} \delta g_{rr}^{- {1 \over
2}}\right]_{r_\infty} \\ \label{vargamma} & = & {A\over4} \delta
\beta_{Hm}^{-1} + {\delta  M_{tot} \over 2}. \end{eqnarray} The
evaluation of the right hand side of the above equation is
straightforward using the proper length variable $\xi$. In Eq.\
(\ref{vargamma}), $\lambda (0)$ is given by Eq.\ (\ref{matemp}) and
$ \lambda(r_\infty)={M_{tot}/r^2_\infty}$ is evaluated using Eq.\
(\ref{deflambda}).  Inserting Eq.\ (\ref{vargamma}) in the
variation of Eq.\ (\ref{smarr2}), gives the differential mass
formula \cite{bch} in the form \cite{ce,ce1}: \begin{equation}
\label{dmff} \delta M_{tot}=\beta_{Hm}^{-1}{\delta A\over 4} +
\delta_{ext} H^\prime_{matter}. \end{equation} Here $ \delta_{ext}
H^\prime_{matter}$ results from the variation of all parameters in
the matter Hamiltonian outside the horizon. Using now Eq.
(\ref{dmff}), and taking the variation with respect to $\beta $ with
the other external parameters kept fixed, we find \begin{equation}
S= \beta^2 \left.  {\delta F \over \delta \beta} \right|_{\mu ,
ext} =- \beta^2 {A \over 4} \left.  {\delta \beta^{-1}_{Hm} \over
\delta \beta} \right|_{\mu}= (1-4 \mu) {A \over 4}. \end{equation}
We find that the entropy of the black hole, as in the absence of
classical surrounding matter, is given by Eq.\ (\ref{entropy}). The
generality of this result shows that the area entropy in presence
of a string instanton is independent of the classical matter
content outside the black hole.  This result  is consistent with
the fact that there is no tree level contribution to the entropy of
the matter outside the black hole. Eq.\ (\ref {state})
generalizes to \begin{equation} \label {fullstate} \beta^{-1} S =
\beta_{Hm}^{-1} {A\over 4}. \end{equation} By performing a
Legendre transform, we find, as before, that the energy of the
solution is unchanged by the presence of the instanton:
\begin{equation} E=F+\beta^{-1}S= M_{tot}. \end{equation}

\section{The Significance of the Euclidean Saddle}
\label{euclideansaddle}

As stated in the Introduction, the canonical partition function
summed over all black hole configurations diverges if the
exponential of the area entropy is identified with the degeneracy
of the corresponding  black hole background. The elimination of the
divergence by restricting the summation to the saddle contribution
of the Euclidean action is not an {\it a priori\ } legitimate procedure.
In this Section we shall first give a thermodynamic justification of
the saddle prescription.  We will then recall \cite{haw2} how the
instability arising in the canonical ensemble approach is avoided
in the microcanonical ensemble. It will then be clear that black
holes of the same mass but with different    Euclidean  sections
have to be considered as  thermodynamically distinct.

We first prove that the saddle prescription follows from
thermodynamics if we assume that $M_{tot}$ represents the
thermodynamic energy and   that the black
hole generates entropy. Namely, we show that these assumptions
require the introduction of an external parameter in the action
(the string tension) when the temperature departs from the value
$\beta^{-1}_{Hm}$ and that the Euclidean periodicity must then be
constrained by adjusting the Euclidean background to be a solution
of the Euclidean equations of motion.

We start from the differential mass formula Eq. (\ref{dmff}):
 \begin{equation} \label{dmf} \delta M_{tot}=\beta_{Hm}^{-1}{\delta
A\over 4} + \delta_{ext} H^\prime_{matter}. \end{equation} It is
important to realize that the derivation of this identity involves
only classical physics, is completely independent of any boundary
terms which were added to the Einstein-Hilbert action, and does not
appeal to Euclidean continuation.

If the system can be thermalized at an  arbitrary temperature
$\beta^{-1}$, the second term in the last equation can be rewritten
as  \begin{equation} \label{micromega} \delta_{ext}
H^\prime_{matter}=\beta^{-1}\delta S_{matter}
+\sum_i\partial_{\lambda_i}H^\prime_{matter}\delta\lambda_i.
\end{equation} The last term is the work done by the variation of
the macroscopic parameters $\lambda_i$ in the action. The first
term, where   $S_{matter}$ is the matter entropy, would arise if
some averaging over ``microscopic'' parameters had been
performed, and is included here for the sake of generality.  Notice that
  the volume   is not a parameter but a dynamical quantity
determined from the equation of motion. Inserting Eq.\
(\ref{micromega}) back into Eq.\ (\ref{dmf}) we get
\begin{equation} \label{ecasbch} \delta
M_{tot}=\beta_{Hm}^{-1}{\delta A\over 4} + \beta^{-1}\delta
S_{matter}
+\sum_i\partial_{\lambda_i}H^\prime_{matter}\delta\lambda_i.
\end{equation} In order to interpret Eq.\ (\ref{ecasbch}) as an
expression of the first principle of thermodynamics we   need the
Bekenstein assumption \cite{bek} that the black hole contributes to
the entropy.  This assumption  forces us to rewrite the first term
as \begin{equation} \label{bhasb} \beta_{Hm}^{-1}{\delta A\over
4}=\beta^{-1}\delta S_{bh}+\sum_iX_i\delta\lambda_i \end{equation}
where $X_i$ are generalized forces. Writing the temperature  as
$\beta=\beta_{Hm}(1-\eta)$, one gets
\begin{equation} \label{dd} \beta_{Hm}^{-1}{\delta A\over 4}=
\beta^{-1}\delta S_{bh} +\beta^{-1}{A\over 4}\delta\eta,
\end{equation} and \begin{equation} \delta
S_{bh}=\delta\left({A\over 4}(1-\eta)\right). \end{equation}
Comparing Eq.\ (\ref{bhasb}) and
Eq.\  (\ref{dd}),  we learn that $\eta$ must be an external
parameter contained in the list $\{\lambda_i\}$. The inverse
temperature $\beta$ is equal to the period of the Euclidean
continuation and $2\pi \eta$ is therefore the deficit angle of the
Euclidean manifold. Since
the integral mass formula is valid independently of the value of
the deficit angle, we must have $\partial_{\eta}H^\prime =0$.
Had we not introduced a string term in the action we would not have been
able to include $\eta$ in the set  $\{\lambda_i\}$. The first
principle would remain valid only at one value of the inverse
temperature namely $\beta=\beta_{Hm}$. In this particular case we
recover of course all the known results. If, however, we introduce
the string, $\eta$ must be a function of the string tension $\mu$.

We now show that thermodynamics imposes that this function be fixed
by the Euclidean equations of motion. It is obvious that the static
matter outside the horizon satisfies the Euclidean equations. What
remains to be proven is that these are also verified at the
Euclidean continuation of the horizon. We stress once more that
Eq.\ (\ref{ecasbch}) is classical and that each term must be
independent of $\hbar$. Since the temperature has  been seen to be
proportional to $\hbar$, it follows, as expected, that $\delta
S_{matter}$ vanishes along the space of classical solutions.
Thus if the integration constant of the entropy is zero, the free
energy of the system is  \begin{eqnarray} \nonumber F&=&
M_{tot}-\beta^{-1}S_{bh}\\ \label{finalfree} &=&
M_{tot}-\beta_{Hm}^{-1}{A\over 4}. \end{eqnarray} The same value
for the free energy Eq.\ (\ref{fullfree}) was deduced from the
saddle point evaluation of the Euclidean action thereby fixing the
value $\eta=4\mu$ as in Eq.\ (\ref{deficit}).  If we had evaluated
the free energy off-shell, we would not have been able to express
the free energy in terms of only surface integrals and we would
not have recovered the result
Eq.\ (\ref{finalfree}). Note also that the value of the generalized
force $\beta^{-1}A/4$ appearing in Eq.\ (\ref{dd}) coincides with
the one in Eq.\ (\ref{dfdmu})  obtained from the functional
integral and is not affected by the surrounding matter.

We now examine the consistency of the thermodynamics arising  from
the Euclidean saddle  with the statistical interpretation in the
microcanonical ensemble. Consider a system composed of gravity and
matter fields. The classical solutions of the equations of motion
define backgrounds for matter and gravitational quantum
fluctuations. We shall neglect the latter and the backreaction of
the   matter field fluctuations on the geometry. We   consider,
within a finite volume $V$, spherically symmetric backgrounds with
a black hole of mass $M$ dressed by a string instanton.   The total
entropy is a function of the total energy $E$ and of $V$. We write
the total Hamiltonian as  $H= H_g +
H^{\prime}_{matter}+H^{f}_{matter}$ where  $H_g$ is the Hamiltonian
of gravity and $H^{f}_{matter}$ is the Hamiltonian characterizing
the fluctuating matter.  Using Eq.\ (\ref{entropy}) the total
entropy  $S_{tot}$ of the system  can be expressed as
\begin{eqnarray} \exp S_{tot}&=&  {\mathop{\rm Tr}} \delta (E-H) \nonumber \\
\label{macro}
 &=& \int_0^{+\infty} dM  \int_{-\infty}^{+\infty} dt\, \exp [ it\,
E + 4\pi M^2(1-4 \mu) -it\, M_{tot} ]  \nonumber  \\
 &  & {\mathop{\rm Tr}}_{f} \exp[{-it H^{f}_{matter}}] \\ \nonumber
 &=&\int_0^{+\infty} dM  \int_{-\infty}^{+\infty} dt\, \exp [ it\, E
 + 4\pi M^2(1-4 \mu) -it M_{tot} -it F_{f}(it)]    \end{eqnarray}
Here, the trace over  matter states  defining the free energy
$F_{f}$ is taken in a background characterized by the black hole
mass $M$ and the on-shell mass value $M_{tot}$ of the Hamiltonian $
H_g + H^{\prime}_{matter}$. Evaluating the $t$-integral by the
steepest descend method gives \begin{eqnarray} \exp S_{tot} &=&
\int_0^{+\infty} dM \exp [ \beta_f E
 + 4\pi M^2(1-4 \mu) -\beta_f M_{tot} -\beta_f F_{f}(\beta_f)]
\nonumber \\ \label{steepest}
 &=& \int_0^{+\infty} dM \exp [
 4\pi M^2(1-4 \mu)  + S_{f}],   \end{eqnarray} where $\beta_f$ is
determined by the equation $E-M_{tot}-\partial_{\beta_f} (\beta_f
F_{f})=0$. $S_{f}$ is the entropy of the fluctuating matter in the
thermodynamic limit of large $E$ and $E-M_{tot}$ of order $E$. The
saddle is a good one because the specific heat of the fluctuating
matter is positive. $S_{tot}$ is finally evaluated  by taking in
Eq.\ (\ref{steepest}) the saddle in the $M$-integral. The saddle
  condition is:
\begin{equation}
{\partial S_{f} \over \partial M_{tot}}
{\partial M_{tot} \over \partial M}+ 8\pi M (1-4\mu)=0.
 \end{equation}
Equivalently,
\begin{equation}
\label{Msaddle}
 {\partial S_{f} \over \partial M_{tot}}
8\pi M \beta_{Hm}^{-1} + 8\pi M (1-4\mu)=0,
\end{equation}
where we used in Eq.\ (\ref{Msaddle}) the differential mass
formula Eq.\ (\ref {dmf}). The Eq.\ (\ref{Msaddle}) gives the equal
temperature condition
\begin{equation}
\label{etemp}
-{\partial S_{f} \over \partial M_{tot}} = \beta_f =\beta_{Hm}
(1-4\mu).
\end {equation}
 One easily verifies,
when $ H^{f}_{matter}$ describes  a relativistic gas in absence
of classical surrounding matter, that the saddle over $M$ is a
maximum when the Hawking condition \cite{haw2}, $ M>(4/5) E$, is
satisfied, independently of the value of $\mu$. This saddle
approximation is thus valid in the thermodynamic limit and the
pathology of the canonical ensemble is avoided. We surmise that
this remains true in general    for regular distribution of matter.

Up to now, we have only considered string instantons which wrap once around
the Euclidean 2-sphere $r=2M$.  The saddle point argument in the
microcanonical ensemble gives back the temperature obtained from the saddle
prescription in the canonical partition function. This confirms that the
configuration with a string instanton has a well defined temperature
different from  the Schwarzschild  black hole in
the absence of string.  One should consequently treat
it as a distinct thermodynamical configuration with a lower entropy and
higher temperature.  One may also generate multi-wrappings. These come
about by mapping the world sheet, which was taken to have the topology of
the sphere, onto an $n$-fold covering of the 2-sphere $r=2M$. Again, their
temperatures, as given by the saddle point in the microcanonical ensemble,
are distinct and define distinct thermodynamical configurations.
It is easily verified that the entropy associated to an $n$-fold covering
is
\begin{equation}
\label{multiwrap}
S_n=(A/4)(1-4n\mu).
\end{equation}
At this stage however, one should question the physical relevance of the
latter result.  Although these solutions are  legitimate for free strings
introduced in our action Eq.\ (\ref{action}), such $n$-fold coverings is
expected to render the result Eq.\ (\ref{multiwrap}) sensitive to
interactions.

\section{Quantum Matter on the Horizon} \label{oneloop}

The  string instanton   at $r=2M$ in Euclidean space does not alter
the classical Lorentzian black hole background  which remains
regular on the horizon.   However dramatic   effects occur  at the
quantum level. To illustrate these we consider the toy model
consisting of the $s$-wave component of a free scalar field
propagating on the Schwarzschild  geometry and we neglect the residual
relativistic potential barrier. This amounts to consider a
2-dimensional scalar field  propagating on the radial subspace of
the 4-geometry. We shall see that   the string instanton induces,
in an inertial frame,   a singularity  in the  vacuum expectation
value of the scalar field energy-momentum tensor  on the horizon.

We parametrize the 2d-background   using the light-like tortoise
coordinates defined by \begin{equation} \label{lightortoise}
u=t-r^\star , \qquad v=t+r^\star, \end{equation} where
\begin{equation} \label{tortoise} dr= (1- {2M \over r}) dr^\star
\quad , \quad r^\star =-\infty \quad  \mbox{at} \quad r=2M,
\end{equation} and \begin{eqnarray} \label{metrictor} ds^2 & = &
-\exp(2 \rho)\  du dv,  \\ \exp(2 \rho) & = & (1-{2M \over r}).
\end{eqnarray}

We compute now the  expectation value of the energy-momentum tensor
of the scalar field using the trace anomaly \cite{bd,dfu}. Note
that in the 2-dimensional approximation the  energy-momentum tensor
of the quantum field $T_{\mu \nu}^{(2)}$ is related to the
4-dimensional one by
 $T_{\mu \nu}^{(2)}=4 \pi r^2 T_{\mu \nu}$.  Since in what follows,
all the quantities are two-dimensional we shall drop   the
superscript $(2)$.

The trace anomaly equation $\left< T \right> =- R/24 \pi$ yields in
the metric Eq. (\ref{metrictor}) \begin{equation} \label{tracet}
\left< T_{uv} \right>=-{1 \over 12 \pi} \partial_u \partial_v \rho.
\end{equation} From Eq. (\ref{tracet}) and from the conservation law
\begin{equation} \label{conserved} T^\mu_{\nu ;\mu}=0 \end{equation}
one gets \begin{eqnarray} \label{tuu} \left< T_{uu} \right> & = &
{1 \over 12 \pi} \left[ -{M \over 2r^3}(1-{2M \over r})-{M^2 \over
4 r^4} \right] + t_u(u)\\  \label{tvv} \left< T_{vv} \right> & = &
{1 \over 12 \pi} \left[ -{M \over 2r^3}(1-{2M\over r})-{M^2 \over 4
r^4} \right] + t_v(v)  \end{eqnarray} where $ t_u(u)$ and $t_v(v)$
are respectively  arbitrary functions of $u$ and $v$.

The usual Hartle-Hawking vacuum is characterized by  the regularity
of $\left<T_{\mu \nu}\right>$ on the past and future horizons in a
local inertial frame. The latter can be parametrized by Kruskal
coordinates $U$ and $V$ defined by \begin{eqnarray} \label{U} U &=&
-4M\lambda e^{-{u\over 4M}} \\ \label{V} V &=& 4M \lambda^{-1}
e^{{v\over 4M}}, \end{eqnarray} where $\lambda$ is an arbitrary
boost parameter and $U=0$ and $V=0$ correspond respectively to the
future horizon and the past horizon. Regularity in the $(U,V)$
frames implies that $\left<T_{uu}\right>$ and $\left<T_{vv}\right>$
vanish on these   horizons like $O\left((1-{2M \over r})^2\right)$.
This condition determines the arbitrary functions   $ t_u(u)$ and
$t_v(v)$ to be constants and equal to \begin{equation} \label{flux}
t_u=t_v={\pi \over 12} {1 \over (8 \pi M)^2}. \end{equation} The
$t_u$ term represents from Eq. (\ref{tuu}) the outgoing flux as $r
\rightarrow \infty$ and the $t_v$ term represents from Eq.
(\ref{tvv}) the ingoing flux as  $r \rightarrow \infty$. Both of
them together correspond to   thermal equilibrium with the
background  at the Hawking temperature $\beta_H = 8 \pi M$.

When the string instanton is present, the temperature gets modified
according to  Eq. (\ref{temp}). We may then characterize   the new
vacuum by taking  \begin{equation} \label{mflux} t_u=t_v={\pi \over
12} {1 \over \left(8 \pi M(1-4\mu) \right)^2}. \end{equation} Using
Eqs. (\ref{tuu}), (\ref{tvv}), (\ref{mflux}) and the coordinate
transformation Eqs. (\ref{U}), (\ref{V}), we see that, in the local
inertial frames, the components of the expectation values of the
energy-momentum have singular parts on the horizons. They   are
given by
 \begin{eqnarray} \label{TUU} \left<T^{sing}_{UU}\right> & = & {\mu
\over 6 \pi U^2}\ {1 - 2 \mu  \over (1-4 \mu)^2}\\ \label{TVV}
\left<T^{sing}_{VV}\right> & = & {\mu \over 6 \pi V^2}\  {1 - 2
\mu  \over (1-4 \mu)^2}. \end{eqnarray}

Eqs.\  (\ref{TUU}) and (\ref{TVV}) translate the fact that
singularities occur on both horizons   when thermal equilibrium in
the Schwarzschild  background is realized at a temperature different from the
Hawking one. This feature  should persist in more realistic four
dimensional computations. Thus the string instanton amounts to  a
materialization of the horizon at the quantum level.    Such a
materialization  constitute a non local effect whose origin can be
traced to the non locality of the string.

The material horizon appears here as a property of quantum fields in
thermal equilibrium in a black hole background. However a material future
horizon could also emerge from the thermal radiation emitted from an
incipient black hole. The point is that in four dimensions, the centrifugal
barrier maintains thermal equilibrium in the vicinity of the horizon
\cite{susshot} and the Hawking radiation may be viewed, even in the
conventional derivation, as a tiny disturbance from equilibrium due to the
escape of the low angular momentum quanta out of the centrifugal
barrier. This departure from thermal equilibrium becomes, close to the
horizon, vanishingly small. Indeed, there, the local temperature tends to
infinity and the fraction of angular momentum modes able to overcome the
barrier becomes of zero measure \cite{emp}. Of course in the original
derivation from local field theory, the global temperature describing this
local equilibrium has the Hawking value Eq.\ (\ref{mass}) and no
singularity will appears on the future horizon. But the huge energy
fluctuations in vacuum close to the horizon renders local field theory
vulnerable to Planckian quantum gravity effects.  Taming these fluctuations
by back reaction may require new elements such as those contained in the
string theory approach and this might lead to materialization of the
horizon.  Such a phenomenon could have important consequences for solving
the unitarity problem posed by black hole evaporation. In fact if unitarity
is to be preserved in our universe, as suggested by 't~Hooft \cite{thooft},
some kind of materialization appears unavoidable
\cite{thooft,susshot,thooft1,weak}. The fact that the materialization
mechanism encountered here is accompanied by a decrease of the area entropy
of the black hole suggests indeed that it is operative in retrieving the
information from the evaporating black hole.

\bigskip

\noindent{\bf Acknowledgements}

We would like to thank  R.~Argurio, R.~Brout and Ph.~Spindel for valuable
discussions. This work was supported in part by the Centre National de la
Recherche Scientifique and the EC Science grant ERB 4050PL920982.

\end{document}